\documentstyle[prl,aps,multicol,epsf]{revtex}

\tighten
\narrowtext
\begin{document}
\draft

\title{Correlation Functions for an Elastic String in a Random
Potential: Instanton Approach}

\author{Ya.~M.~Blanter$^{a,b}$ and V.~M.~Vinokur$^{b}$}
\address{
$^a$ Department of Applied Physics and DIMES, Delft University of
Technology, Lorentzweg 1, 2628 CJ Delft, The Netherlands\\
$^b$ Materials Science Division, Argonne National Laboratory, Argonne,
IL 60439 \\}
\date{\today}
\maketitle
\begin{abstract}
We develop an instanton technique for calculations of correlation
functions characterizing statistical behavior of the elastic string in
disordered media and apply the proposed approach to correlations of
string free energies corresponding to different low-lying metastable
positions. We find high-energy tails of correlation functions for
the case of long-range disorder (the disorder correlation length well
exceeds the characteristic distance between the sequential string
positions) and short-range disorder with the correlation length much
smaller then the characteristic string displacements. The former case
refers to energy distributions and correlations on the distances below
the Larkin correlation length, while the latter describes correlations
on the large spatial scales relevant for the creep dynamics.
\end{abstract}

\pacs{PACS numbers: 61.41.+e, 74.60.Ge, 05.40.-a}

\begin{multicols}{2}

An elastic string in two-dimensional random potential is an
archetypal problem of statistical physics, with applications to a
widest variety of systems and phenomena. The examples include vortices
in superconductors \cite{htc}, dislocations \cite{disl}, domain walls 
\cite{domainwalls}, and non-Hermitian quantum mechanics
\cite{Localiz}. The string that energetically disfavors overhangs (like
vortices) maps onto {\em directed polymer} \cite{HHZ}. In the random
potential the string bends to adjust itself to the pinning potential
relief. Directed polymer approximation corresponds to {\em weak
pinning}, where the string shape is smooth and coincides with the path
of the classical particle traveling through the corresponding rugged
energy landscape. The Hamiltonian of the string has the form
\cite{Huse}  
\begin{equation} \label{hamilt}
H = \int dx \left[ \frac{\kappa}{2} \left( \frac{d\zeta}{dx}
\right)^2 + V(x,\zeta) \right], \ \ \ \ \left\vert \frac{\partial
\zeta}{\partial x} \right\vert \ll 1.
\end{equation}
Here $x$ is the coordinate along the preferential direction (for
small displacements it also measures the length of the segment),
$\zeta(x)$ is transverse displacement, $\kappa$ is the elastic
constant, and the random potential $V(x,\zeta)$ is routinely assumed
to be Gaussian, with zero average $\langle V \rangle = 0$ and the
correlation function
\begin{displaymath}
\left\langle V(x,\zeta) V(x',
\zeta') \right\rangle = \beta \delta(x-x') K(\zeta - \zeta'), \
\int K(y) dy = 1.
\end{displaymath}
The amplitude $\beta$ depends on the type of disorder. For instance,
for a vortex pinned in a plane by point defects, $\beta \approx
\xi_0^2 n_i V^2$, with $\xi_0$, $n_i$, and $V$ being the core radius,
concentration of defects, and the energy of vortex-defect interaction,
respectively. 

Much of the early effort was concentrated on the fluctuations of the
free energy of the string. Let the left end of the 
string be fixed at $\zeta = 0$, $x_L \to -\infty$. The quantity of
interest is then the free energy of the string $\varepsilon (x,
\zeta)$ as the function of the position of its right end $(x_L+x,
\zeta)$. Knowledge of the mean free energy $\langle \varepsilon
\rangle$ is important for calculation of dynamical quantities,
such as pinning energy or drift velocity. Various tools, which, among
others, include numerical analysis \cite{Huse}, Bethe Ansatz 
solution \cite{KN}, and power counting based on the effective action
\cite{Ioffe}, yield the result for the disorder-averaged free energy
$\langle \varepsilon \rangle \propto x^{1/3}$ and the average
displacement of the string $\langle \zeta \rangle \propto x^{2/3}$. 

The dynamic response, however, requires more detailed statistical
description, since in such a complex system a non-self-averaging
behavior can be expected. Many of the results have been already made
available in the course of 15-year-long research, see
Ref. \onlinecite{HHZ} for an excellent review. For recent
developments, we cite calculations of the free energy cumulants
\cite{Brunet} and tails of its distribution function \cite{Gorokhov}.  

So far, attention has been mainly drawn to statistical properties
of the free energy for a fixed position of the end,
$\varepsilon(x,\zeta)$. Of interest are also correlation
functions describing different positions of the string
important for its slow dynamics (noise, velocity
correlations). Conceptually, they also test non-Gaussian shape of
distributions. In this Letter, we make the first step in this
direction and evaluate the distribution function,   
\begin{equation} \label{defin}
P(u) = \left\langle \delta \left[ u - \varepsilon (0,\zeta_1) +
\varepsilon (0, \zeta_2) \right] \right\rangle,
\end{equation}
which describes correlation between the sequential energies of the
string as it is displaced transversely to the preferential
direction. In what follows, we assume $\zeta_1 \ge \zeta_2$.  

{\bf Model}. The starting point of our theory is the equation for the
free energy $\varepsilon$ which can be derived by the transfer matrix
method from Eq. (\ref{hamilt}) and reads 
\begin{equation} \label{eqbasic}
\frac{\partial \varepsilon}{\partial x} + \frac{1}{2\kappa} \left(
\frac{\partial \varepsilon}{\partial \zeta} \right)^2 -
\frac{T}{2\kappa} \frac{\partial^2 \varepsilon}{\partial \zeta^2} =
V(x,\zeta),
\end{equation}
where $T$ is temperature. Eq. (\ref{eqbasic}) is easily recognized
as Kardar-Parisi-Zhang (KPZ) equation in $1+1$ dimensions
\cite{HHZ}. To this end, $x$ and $\zeta$ can be identified as time and
coordinate, respectively. 

We focus on the behavior of the correlation function
(\ref{defin}) at large $u$. In this case, $P(u)$ is the probability
that the energy changes considerably upon small displacement of the
string. This probability is expected to be exponentially small, and an
appropriate technique to calculate exponential tails is the instanton
approach, allowing us to avoid difficulties associated with the
replica method. The Laplace transform of the function (\ref{defin}) can
be presented as the functional integral \cite{Ioffe}, 
\begin{equation} \label{laplace1}
\Pi (\lambda) = \left\langle e^{\lambda [\varepsilon(0,\zeta_1) -
\varepsilon(0,\zeta_2)]} \right\rangle = \int D\varepsilon D\psi e^{S
[\varepsilon, \psi]},
\end{equation}
with the effective action
\begin{eqnarray} \label{action0}
S[\varepsilon, \psi] & = & i \int dx d\zeta \psi(x, \zeta) \left[
\frac{\partial \varepsilon}{\partial x} + \frac{1}{2\kappa} \left(
\frac{\partial \varepsilon}{\partial \zeta} \right)^2
- \frac{T}{2\kappa} \frac{\partial^2 \varepsilon}{\partial \zeta^2}
\right] \nonumber \\ & - & \beta \int dx d\zeta d\zeta' \psi(x,\zeta)
K(\zeta - \zeta') \psi(x, \zeta') \nonumber \\
& + & \lambda [\varepsilon(0,\zeta_1) - \varepsilon(0,\zeta_2)].  
\end{eqnarray}
To arrive to Eq. (\ref{action0}), we have used the identity
\begin{displaymath}
\int D\varepsilon \delta \left[ \frac{\partial \varepsilon}{\partial
x} + \frac{1}{2\kappa} \left( \frac{\partial \varepsilon}{\partial
\zeta} \right)^2  - \frac{T}{2\kappa} \frac{\partial^2
\varepsilon}{\partial \zeta^2} - V \right] = const,
\end{displaymath}
which is a consequence of causality, and subsequently performed
averaging over the random potential $V$. 

The field theory (\ref{action0}) is essentially two-dimensional. To
move further, we extend the technique developed by Gurarie and Migdal
\cite{Gurarie} (GM), who studied velocity correlations in the Burgers
equation. This is, in order to describe the tails of the correlation
function (\ref{defin}) we have to find the instanton (saddle-point)
trajectory and calculate the action $S_{in}$ at this trajectory. The
result for $\Pi (\lambda)$ with the exponential accuracy reads  
\begin{equation} \label{laplace2}
\Pi (\lambda) = \exp \left[ S_{in} (\lambda) - S_{in} (0) \right].
\end{equation}
Prefactors can be obtained systematically by expanding the action
around the instanton path, but this goes beyond the scope of this
Letter. 

The saddle-point equations for the effective action $S[\varepsilon,
\psi]$ have the form
\end{multicols}
\widetext
\hrulefill\hfill\
\begin{eqnarray}
& & \frac{\partial \varepsilon}{\partial x} + \frac{1}{2\kappa}
\left( \frac{\partial \varepsilon}{\partial \zeta} \right)^2 -
\frac{T}{2\kappa} \frac{\partial^2 \epsilon}{\partial \zeta^2} =
-2i\beta \int d\zeta' K(\zeta - \zeta') \psi(x, \zeta');
\label{saddle1} \\
& & \frac{\partial \psi}{\partial x} + \frac{1}{\kappa}
\frac{\partial}{\partial \zeta} \left( \psi \frac{\partial
\varepsilon}{\partial \zeta} \right) + \frac{T}{2\kappa}
\frac{\partial^2 \psi}{\partial \zeta^2} = -i\lambda \delta(x) \left[
\delta(\zeta - \zeta_1) - \delta(\zeta - \zeta_2) \right].
\label{saddle2}
\end{eqnarray}
\hfill\hrulefill
\begin{multicols}{2}

{\bf Long-range disorder}. We assume first that the function $K$ is
only slightly changed on the scale of $\zeta_1 - \zeta_2$. It can be
then expanded, $K(y) = k_0 - k_1 y^2/2$. In the subsequent analysis,
we follow GM. 

Note that Eq. (\ref{saddle2}) can only have non-zero solutions for $x
< 0$ (``diffusion in reverse time''). The field $\psi$ is
discontinuous at $x=0$; integrating Eq. (\ref{saddle2}) between $-0$
and $+0$, we obtain the boundary condition, 
\begin{equation} \label{cond1}
\psi(x=-0, \zeta) = i\lambda \left[ \delta(\zeta - \zeta_1) -
\delta(\zeta - \zeta_2) \right].
\end{equation}

To solve the saddle-point equations, we disregard the term with
$\partial^2 \psi/\partial \zeta^2$ in Eq. (\ref{saddle2}) for a while.   
Then, due to Eq. (\ref{cond1}), the function $\psi$ always remains the
sum of two delta-functions,  
\begin{equation} \label{try1}
\tilde \psi (x, \zeta) = i \mu (x) \left[ \delta(\zeta - \rho_1 (x)) -
\delta(\zeta - \rho_2 (x)) \right], 
\end{equation}
where the tilde means that this solves the ``incomplete''
equation. Eq. (\ref{cond1}) implies $\mu(0) = \lambda$, $\rho_{1,2}
(0) = \zeta_{1,2}$. From Eq. (\ref{saddle1}) we then find $\varepsilon
(x, \zeta) = \tilde a(x) + \tilde b(x) \zeta$, with  
\begin{eqnarray*}
\tilde a' + \tilde b^2/2\kappa = -\beta k_1 \mu (\rho_1^2 -
\rho_2^2); \ \  
\tilde b' = 2\beta k_1 \mu (\rho_1 - \rho_2).
\end{eqnarray*}
Substituting this back into Eq. (\ref{saddle2}), we find $\mu(x) =
\lambda$ and $\rho_1 (x) - \rho_2 (x) = \zeta_1 - \zeta_2$. These 
solutions and, consequently, the resulting form of the distribution
function, are very different from those for Burgers equation
\cite{Gurarie}.  

Now we include the term with $\partial^2 \psi/\partial \zeta^2$ into
consideration. We look for a solution of Eq. (\ref{saddle1}) in the
form of a linear function of $\zeta$, $\varepsilon = a(x) + b(x)
\zeta$ (the consistency is checked afterwords). The {\em linear}
equation (\ref{saddle2}) is easily solved and yields  
\begin{eqnarray} \label{try2}
\psi(x, \zeta) & = & \frac{i\lambda \kappa^{1/2}}{(2\pi T \vert x
\vert )^{1/2}} \left[ e^{\frac{\kappa (\zeta - \rho_1)}{2Tx}} -
e^{\frac{\kappa (\zeta - \rho_2)}{2Tx}} \right], 
\end{eqnarray}
with $\rho_1' = \rho_2' = b(x)/\kappa$. Substituting this into
rhs of Eq. (\ref{saddle1}), we find that the latter is
temperature independent, {\em i.e.} it remains the same as that
without the diffusion term. Thus, the linear Ansatz for $\varepsilon$
is consistent. Also, we find $b' = 2\beta \kappa_1 \lambda (\rho_1 -
\rho_2)$. The solution is then $\rho_{1,2} (x) = \zeta_{1,2} + b_0
x$.  

Next, we calculate the action $S$ along the instanton
trajectory. As seen from the saddle-point equations, the first term in 
Eq. (\ref{action0}) is just doubled the second one. Multiplying
Eq. (\ref{saddle2}) by $\varepsilon$,  integrating it over $x$ 
(from $-\infty$ to $-0$) and over $\zeta$, and comparing with
Eq. (\ref{saddle1}), we find   
\begin{displaymath}
\lambda [\varepsilon(0, \zeta_1) - \varepsilon(0, \zeta_2)] =
-2\beta \int \psi K \psi - \frac{i}{2\kappa} \int dx d\zeta \psi
\left( \frac{\partial \varepsilon}{\partial \zeta} \right)^2. 
\end{displaymath}    
Since $\varepsilon$ depends on $\zeta$ linearly, and $\int \psi d\zeta
= 0$, the last term in the right-hand side vanishes. The instanton
action acquires the form
\begin{equation} \label{action1}
S_{in} = -\beta \int_{-\infty}^{-0} dx \int d\zeta d\zeta'
\psi(x, \zeta) K(\zeta - \zeta') \psi(x, \zeta').
\end{equation}
Substituting the instanton solution (\ref{try2}) into
Eq. (\ref{action1}), we find that the term 
proportional to $k_0$ in the effective action vanishes, while the
contribution with $k_1$ diverges in the limit of large negative
$x$. This is because our consideration is limited to distances shorter
than the correlation length of the random potential $\xi \sim
(k_0/k_1)^{1/2}$, so that we could replace $K(y)$ by its
expansion. This cut-off effectively defines the constant
$b_0$. Replacing integrals over $\zeta$ in Eq. (\ref{action1}) in
infinite limits by integrals from $-\xi/2$ to $\xi/2$ and calculating
them, we arrive at the cut-off for $x$,  
\begin{eqnarray} \label{cut0}
x_c = \left\{ \begin{array}{lr} 
-\kappa \xi^2/T,\ \ \  & T > T_c \\
-(\kappa \xi/\beta k_1 \lambda \zeta_0)^{1/2},\ \ \  & T < T_c
\end{array}
\right. \ \ ,
\end{eqnarray}
\vspace{1.5cm}

\begin{figure}    
\narrowtext    
\epsfxsize=9.cm   
\centerline{\epsffile{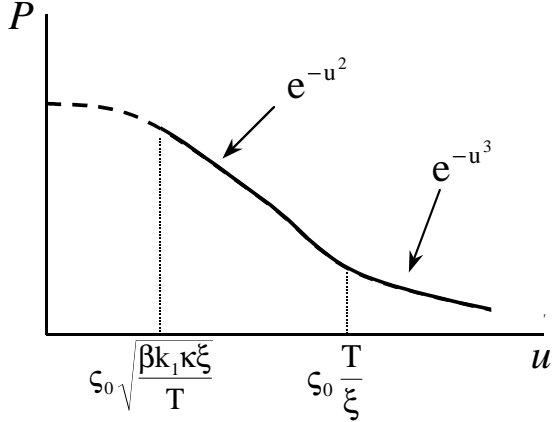}} 
\vspace{0.2cm}   
\caption{Distribution function $P(u)$ for long-ranged correlated
disorder, Eq. (\protect\ref{distribshort}).}    
\label{Fig1}    
\end{figure}    
\noindent with $\zeta_0 = \zeta_1 - \zeta_2 > 0$ and $T_c = (\beta k_1 
\kappa \zeta_0 \xi \lambda)^{1/2}$. The instanton action reads 
$S_{in} = \beta k_1 \zeta_0^2 \lambda^2 \vert x_c \vert$. Note that,
since our cut-off procedure is somewhat arbitrary, we have removed all
numerical factors from the action. Laplace-transforming
Eq.(\ref{laplace2}), we arrive at the expression 
\begin{eqnarray} \label{distribshort}
P(u) = \left\{ \begin{array}{lr} e^{-c'Tu^2/(\beta k_1 \kappa \xi^2
\zeta_0^2})\ , & u \ll T\zeta_0/\xi \\
e^{-cu^3/(\beta k_1 \kappa \xi
\zeta_0^3})\ , & u \gg T\zeta_0/\xi 
\end{array} \right. ,   
\end{eqnarray}
which is illustrated on Fig.~1, with $c \sim c' \sim 1$. 

The result (\ref{distribshort}) was obtained by instanton approach,
which means it has to be exponentially small. In particular, this
requires $u >0$. We thus get the conditions $u 
\gg \zeta_0 (\beta k_1 \kappa \xi)^{1/3}$ for the 
elasticity-controlled regime ($P(u) \propto \exp(-u^3)$) and $\zeta_0
(\beta k_1 \kappa \xi^2/T)^{1/2} \ll u \ll \zeta_0 T/\xi$ for the
temperature-controlled regime ($P( u) \propto \exp(-u^2)$). The latter
result only holds for temperatures higher than $T_d = \xi^{4/3}
(\beta k_1 \kappa)^{1/3}$, which does not depend on $\zeta_0$ and has
a meaning of depinning temperature --- the typical pinning energy
on the Larkin correlation length. Above the depinning temperature
correlation functions decay faster than exponentially \cite{htc}. 

{\bf Short-ranged disorder}. We take now $K(y) =
\delta(y)$. At $x=0$, the field $\psi$ is a set of a delta-peak
(located at $\zeta = \zeta_1$) and a delta-dip ($\zeta = \zeta_2$),
see Eq. (\ref{cond1}). As we trace the evolution in reverse time,
these features smear and move. For short times, before
(if ever) they intersect or smear so much that they become
indistinguishable, $\psi$ remains a peak-dip function, centered
at $\zeta = \rho_1 (x)$ and $\zeta = \rho_2 (x)$, and sharply
vanishing away from these points.  

Let us calculate the energy created by a peak-shaped field $\psi$,
located at $\zeta = \rho$, far from this point. At large distances
from the peak, the solution must have a scaling form. Eq. (\ref{saddle1}) 
with the zero right-hand side allows for only one scaling solution
$\varepsilon = f((\zeta-\rho)/x^{1/2})$ vanishing at $x \to
-\infty$, which at large scales, where the term with the second
derivative can be disregarded, becomes $\varepsilon 
= \kappa(\zeta-\rho)^2/2x$. For $\vert \zeta - \rho \vert <
(Tx/\kappa)^{1/2}$ the diffusion term is important, and this
simple form of the scaling solution does not apply. 

Our solution has an obvious drawback: It diverges as $x$ goes to
zero. To amend this, we write
\begin{equation} \label{try10}
\varepsilon = \frac{\kappa (\zeta - \rho)^2}{2 (x - x_0)}, \ \ \ x_0 >
0. 
\end{equation}    
The regularization constant $x_0$ plays an important role and is found
from the following considerations. For $x=0$, the energy given by
Eq. (\ref{try10}) is $\varepsilon_0 = - \kappa \zeta^2/x_0$. In the
scaling regime, the only one relevant length scale is $\zeta_0 =
\zeta_1 - \zeta_2$, and the only relevant energy scale is
temperature. Therefore, $T \sim \kappa \zeta_0^2 /x_0$, 
whence $x_0 \sim \kappa \zeta_0^2 /T$. The numerical factor remains
undetermined, but can be deduced from numerical solutions of
Eqs. (\ref{saddle1}), (\ref{saddle2}).    

Now we return to Eq. (\ref{saddle2}) and solve it at small
$\vert x \vert$, when the peak and the dip are well separated. We
write $\psi(x, \zeta)$ in the form $\psi_{peak} + \psi_{dip}$. The peak
(located near $\zeta = \rho_1(x)$) sees then the energy produced by
itself and the energy produced by the dip. The energy of an isolated
peak (\ref{try10}) grows as $\zeta^2$ far from the peak, and hence at
the point $\rho_1$ the energy produced by the dip is much greater than
the peak contribution. If, in addition, $\rho_2 - \rho_1 \gg
(T(x_0-x)/\kappa)^{1/2}$, we can use Eq. (\ref{try10}) to evaluate the
coefficient $\partial \varepsilon/\partial \zeta$ in
Eq. (\ref{saddle2}). We get then the equation
\begin{equation} \label{peak1}
\frac{\partial \psi_{peak}}{\partial x} + \frac{\psi_{peak}}{x - x_0}
+ \frac{\rho_1 - \rho_2}{x-x_0} \frac{\partial \psi_{peak}}{\partial
\zeta} + \frac{T}{2\kappa} \frac{\partial^2 \phi}{\partial \zeta^2}
=0,
\end{equation}
with the boundary condition $\psi_{peak} (x=0) = i\lambda \delta(\zeta
- \zeta_1)$. 

Writing a similar equation for the dip ($\zeta \approx \rho_2$) and
solving both equations, we obtain
\begin{eqnarray} \label{peak2}
\left\{ \begin{array}{c} \psi_{peak} \\ \psi_{dip} \end{array}
\right\} & = & \pm \frac{i \lambda x_0 \kappa^{1/2}}{(2\pi T \vert x
\vert)^{1/2}} \frac{1}{x_0 - x} \exp \left( - \frac{\kappa (\zeta -
\rho_{1,2})^2}{2T \vert x \vert} \right), \nonumber \\
\left\{ \begin{array}{c} \rho_1' \\ \rho_2' \end{array}
\right\} & = & \pm \frac{\rho_1 - \rho_2}{x - x_0}.    
\end{eqnarray}  

As follows from Eq. (\ref{peak2}), positions of the peak and the dip
are pulled apart. The distance between them grows as $(x-x_0)^2$; at
the same time, they smear as $\vert x \vert^{1/2}$; thus, they never
overlap and can be considered as well separated {\em at any}
$x$. Using Eq. (\ref{peak2}) to calculate the instanton action, we get 
\begin{equation} \label{inst10}
S_{in} = \tilde a \beta \lambda^2 \frac{\kappa \zeta_0}{T}, \ \ \ \zeta_0
= \zeta_1 - \zeta_2, \ \ \ \tilde a \sim 1, 
\end{equation} 
which translates into the expression for the distribution function 
\begin{equation} \label{inst11}
P(u) = \exp \left( -\frac{a T u^2}{\beta \kappa \zeta_0} \right),
\end{equation}
with a numerical constant $a$ of order one. The expression
(\ref{inst11}) is valid for temperatures much higher than 
$\beta \kappa \zeta_0/u^2$ and is an analog of the upper line of
Eq. (\ref{distribshort}). We conjecture that the long-$u$ tail of
the distribution function (the analog of the lower line of
Eq. (\ref{distribshort})) also exists, though we were not able to
obtain it with this method of solving Eqs. (\ref{saddle1}),
(\ref{saddle2}).  

{\bf Discussion and conclusions}. Eqs. (\ref{distribshort}) and
(\ref{inst11}) constitute the central result of this Letter. The
function $P(u)$ has a meaning of the probability that the difference 
of energies of a string of the same length and the transverse
displacements $\zeta_1$ and $\zeta_2$, equals $u$. An instanton
solution corresponds to the situation when at $\zeta_1$ is a minimum
of the potential energy while at $\zeta_2$ is the {\em neighboring}
maximum. $P[u]$, thus, measures the height of the barrier of the free
energy relief. For large $u$ to find a high barrier is quite
improbable, consequently the result is exponentially small.  

If the string has a finite transverse size $\xi_0$, the correlation
radius $\xi$ of the disordered potential is of order of
$\xi_0$. Eq. (\ref{distribshort}) describes thus the case when the 
string is displaced at a distance small compared with its transverse
dimension, and Eq. (\ref{inst11}) applies in the opposite regime of
long distance. 

These results have an important conceptual value, since they describe
correlation functions of energy at different 
positions for the directed polymer problem. We argue now that they
also provide certain predictions for experimentally observable
dynamical properties of the string. We suggest two types of
experiment, which certainly do not exhaust all the opportunities. 

(i) Creep of domain walls. If a domain wall moves from one position to
another one, the distance between the two positions can be
measured. The quantity $\zeta_0$, which is the distance between a
position and a maximum separating the two, is not known, but can be
determined if the wall travels back and forth. The tunneling rate is
determined by the height of the barrier $u$. A similar information can
be extracted from the random telegraph noise resulting from the motion
of the string between two positions. The time spent in each position
is determined by the barrier to be overcome seen from this
position. These considerations do not take into account a possibility
of macroscopic quantum tunneling, which has to be considered
separately.       

(ii) Distribution of pinning energies. The string is (locally)
depinned if it moves from one free energy minimum to the next one. The
pinning energy corresponds precisely to the quantity $u$
above. However, globally one has to average over the distribution of
the displacements $\zeta_0$, {\em i.e.} to know the distribution
of distances between adjacent minima and maxima of the random
potential. Measurements of the distributions of the pinning energy
compared with our results can provide an information about this
distribution, which, to our knowledge, has not been previously
discussed.  

In conclusion, we have developed an instanton approach to calculations
of various correlation functions describing statistical behavior of
the elastic string in the two-dimensional disordered potential.  We
applied our technique to the investigation of correlations of free
energies corresponding to different low-lying metastable positions of the
string. We have found the asymptotic behavior of such energy-energy
correlations for the moderate spatial scales (within Larkin
correlation length) and the large scales, exceeding Larkin length.
The latter situation corresponds to the conditions of the creep
dynamics. We have discussed applications of our results to the dynamic
response and noise in various two-dimensional systems such as domain
walls, vortices, and dislocations in thin films.  

The work was supported by the U.S. DOE, Office of Science, under
contract No. W31-109-ENG-38.

\end{multicols}

\end{document}